\documentclass[11pt]{gh2013}

\usepackage{mathptmx}       
\usepackage{helvet}         
\usepackage{courier}        
\usepackage{makeidx}         
\usepackage{graphicx}        
\usepackage{multicol}        

\begin{document}

\title*{Observations of the Most Massive Deeply Embedded Star Clusters in the Milky Way}
\titlerunning{Massive Deeply Embedded Star Clusters in the Milky Way}
\author{Roberto Galv\'an-Madrid$^{1}$ and Hauyu Baobab Liu$^{2}$}
\institute{$^{1}$European Southern Observatory, Karl-Schwarzschild-Str. 2, 85748 Garching, Germany, 
\email{rgalvan@eso.org}
\\ $^{2}$ Academia Sinica Institute of Astronomy and Astrophysics, P.O. Box 23-141, 
Taipei 106}
%
%
\maketitle


\vskip -3.5 cm  
\abstract{
We summarize our comprehensive gas surveys of some of the most luminous, deeply embedded 
(optically obscured) star formation regions in the Milky Way, which are the local cases of massive star 
clusters and/or associations in the making. Our approach emphasizes multi-scale, multi-resolution imaging 
in dust and free-free continuum, as well as in molecular- and hydrogen recombination lines, to trace the 
multiple gas components from 0.1 pc (core scale) all the way up to the scales of the entire giant 
molecular cloud (GMC), or $\sim 100$ pc. We highlight our results in W49A, the most luminous Galactic star 
formation region ($L\sim 10^7~L_\odot$), which appears to be forming a young massive cluster 
(or a binary star cluster) 
with $M_\star > 5\times10^4~M_\odot$ that may remain bound after gas dispersal. The surveyed sources share 
elements in common, in particular, the 10-100 pc scale GMCs are filamentary but have one or two central 
condensations (clumps) far denser than the surrounding filaments and that host the (forming) massive stars. 
}

\section{Introduction}

Young massive clusters (YMCs) have stellar masses $M_\mathrm{cl}>10^4$ M$_\odot$, scales of a 
few pc,  and are younger than a $\sim 100$ Myr. They probably represent 
the young end of the so-called super star clusters (SSCs) found in starbursting galaxies 
\cite{Whitmore93,Meurer95}. 
It is also possible that some of them are young analogues of globular clusters (GCs). 
Recent reviews are \cite{PZ10} and \cite{Longmore14}. 

Understanding how YMCs 
form is key to understand the properties of star forming galaxies. 
Some studies have targeted the progenitor giant molecular clouds (GMCs) in nearby extragalactic systems
\cite{Santangelo09,Wei12}. 
The deeply embedded, most luminous ($L>10^6~L_\odot$) 
Galactic star formation regions stand out as the obvious candidates to be
active local YMC formation sites \cite{Carlhoff13}. These objects are rare in our Milky
Way \cite{Ginsburg12}. The evolution of their natal molecular clouds are governed
by the interplay of gravity, turbulence, (proto-) stellar feed
back, and potentially magnetic fields. The open 
questions to answer are: What are the properties and physical conditions
of the molecular clouds that give birth to YMCs? What are the
effects of YMC feedback upon its own parental cloud?

We are carrying a program to study in detail the GMCs of the most luminous 
star formation regions in the Milky Way. We emphasize multi-scale mapping 
using both interferometers and single dishes, both separately and combined. 
Here we highlight our first results in W49A, and compare to our 
previous results.

\section{W49A}
\label{sec:w49a}

W49A 
is the most luminous star formation region in the Milky Way ($L\sim10^{7.2}$ $L_\odot$), 
embedded in one of the most massive giant molecular clouds (GMCs), 
$M_\mathrm{gas}\sim10^6$ $M_\odot$ \cite{Miyawaki09}. 
The GMC has an extent of $l>100$ pc, 
but all the prominent 
star formation resides in the central $\sim 20$ pc, and peaks in the subregion known as W49N.  
Part of the stellar population in W49N is already visible in the near-IR and 
its mass 
has been estimated at $M_\mathrm{cl}\sim4\times10^4$ $M_\odot$ \cite{HomeierAlves05}, 
whereas the part associated with the  most compact HII regions is not 
visible at wavelengths shorter than mid-IR.

We have performed extensive mapping observations of molecular lines toward W49A 
using the Submillimeter Array (SMA), the Purple Mountain Observatory 14m Telescope (PMO-14m), 
and the IRAM-30m Telescope \cite{GM13}. 
The PMO-14m images with $~3$ pc resolution resolved several 10-30 pc scale radially converging gas 
filaments (Figure 1, left). The
most active cluster-forming region W49N coincides with the convergence of these filaments. 
The two less prominent neighbors
W49S and W49SW appear to be formed in the two densest gas filaments connecting to the southeast 
and southwest of W49N. The larger-scale filaments are clumpy.

\begin{figure*}[!h]
  \centering
  \includegraphics[width=\textwidth]{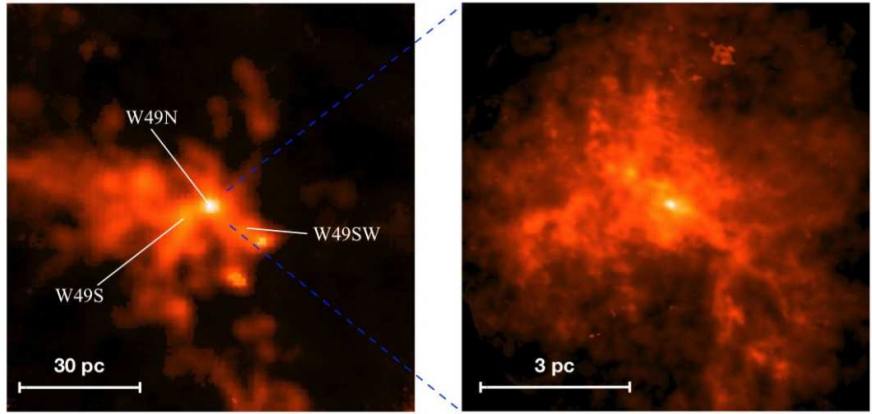}
  \caption{Mass surface density $\Sigma$ maps obtained from CO-isotopologue line ratios. 
The left panel shows the zoomed-out measurement from the PMO-14m telescope CO and
$^{13}$CO 1--0 maps. The right panel shows the zoomed-in measurement from the SMA mosaics 
combined with IRAM-30m telescope maps of $^{13}$CO and C$^{18}$O 2--1. 
The embedded OB cluster-forming regions W49N, W49S, and W49SW are marked \cite{GM13}.}
  \label{fig:f1}
\end{figure*}

The combined IRAM-30m + SMA images with $\sim0.1$ pc resolution
further trace a triple, centrally condensed filamentary structure
that peaks toward the central parsec scale ring of HC HII regions
in W49N, which is known to host dozens of deeply embedded
(maybe still accreting) O-type stars (Figure 1, right). In addition,
localized UC HII regions are also found in individual filaments.
Our finding suggests that the W49A starburst most likely formed
from global gravitational contraction with localized collapse in a
'hub-filament' geometry.

From multi-scale observations of CO isotopologues, we derived
a total molecular mass for the GMC $M_\mathrm{gas} \sim 1.1 \times 10^6~M_\odot$ within a radius 
of 60 pc, and $M_\mathrm{gas} \sim 2 \times 10^5~M_\odot$ within 6 pc \cite{GM13}. Approximately 
$\sim 20\%$ of the gas mass in concentrated in $\sim 0.1\%$ of the volume. The mass reservoir 
is enough to form a YMC as massive as a globular cluster. 
Currently, only $\sim 1\%$ of the gas in the central few pc is ionized, which indicates 
that feedback is still not enough to significantly clear the GMC. 
The resulting stellar content might remain
as a gravitationally bound massive star cluster or a small system
of bound star clusters. Further analysis of this data is ongoing.

\section{Other regions}
\label{sec:other}

Before our MUSCLE program in W49A \cite{GM13} we have observed several other 
optically obscured, very luminous star formation regions to reveal the relation 
between the cloud to clump to core scales \cite{GM09,GM10,Liu12a,Liu12b}. 

Figure 2 shows an early result in G20.08-0.04 ($L \sim 7 \times 10^5~L_\odot$), 
where velocity gradients interpreted 
as rotation are seen in dense gas tracers (NH$_3$ and CH$_3$CN) and infall 
is seen in NH$_3$ absorption against the background HC HII regions too. 
Both rotation and infall are seen at scales from $> 1$ pc (clump) 
to $<0.1$ pc (core), indicating continuity 
in the accretion flow at multiple scales \cite{GM09}. 

Figure 3 shows G10.6--0.4 ($L \sim 9 \times 10^5~L_\odot$). The similarities with W49A 
are striking, although scaled down. Filaments converge radially in a central hub which hosts 
most of the current OB star formation. The overall structure is hierarchical, 
with subfragmentations along the filaments which have further substructure.

\begin{figure*}
  \centering
  \includegraphics[width=\textwidth]{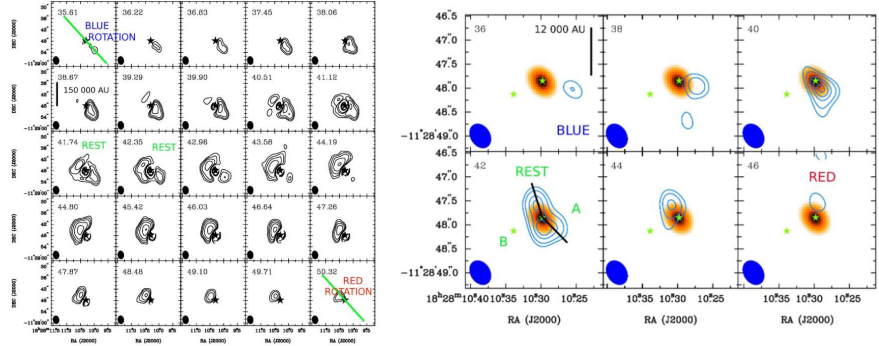}
  \caption{Rotation in the dense gas at multiple scales in G20.08N. The left panel shows the parsec 
scale clump seen in NH$_3$ (3,3) emission. The star marks the position of the central HC HII regions  
against which NH$_3$ is seen in absorption. The right panel zooms into the central 0.1 pc, where 
a similar velocity gradient is seen in denser, warmer gas traced by CH$_3$CN emission, although with less 
specific angular momentum \cite{GM09}.}
  \label{fig:f1}
\end{figure*}

\begin{figure*}
  \centering
  \includegraphics[width=0.6\textwidth]{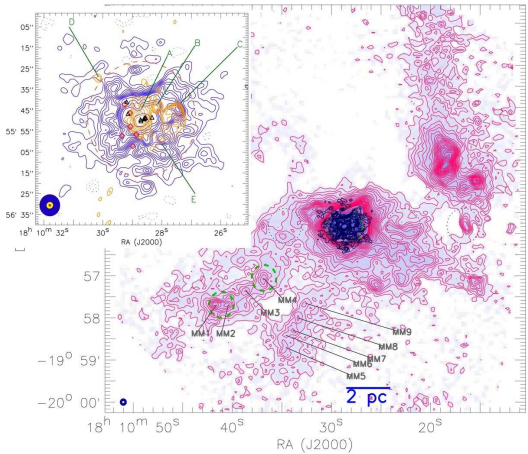}
  \caption{Large-scale mapping of G10.6--0.4. The central panel shows IRAM 30m and SMA 1.2-mm continuum
emission. The inset shows the 30m+SMA mm image (blue contours) enclosing the ionized gas (yellow) 
of the central OB cluster \cite{Liu12a}.}
  \label{fig:f1}
\end{figure*}

\begin{acknowledgement}
RGM and HBL kindly acknowledge the organizers of the Guillermo Haro conference where these 
results were presented. 

\end{acknowledgement}

\end{document}